# Cr doping-induced ferromagnetism in the spin-glass $Cd_{1-x}Mn_x Te$ studied by x-ray magnetic circular dichroism

V. K. Verma,[1,2] S. Sakamoto,[1,3*] K. Ishikawa,[4] V. R. Singh,[1,5] K. Ishigami,[1,6] G. Shibata,[1,7] T. Kadono,[1] T. Koide,[8] S. Kuroda[4] and A. Fujimori[1,9]

[1]*Department of Physics, University of Tokyo, Bunkyo-ku, Tokyo 113-0033, Japan*
[2]*Department of Physics, Madanapalle Institute of Technology & Science, Madanapalle 517325, Andhra Pradesh, India*
[3]*Institute for Solid State Physics, University of Tokyo, Kashiwa, Chiba 277-8581, Japan*
[4]*Institute of Materials Science, University of Tsukuba, Tsukuba, Ibaraki 305-8573, Japan*
[5]*Department of Physics, Central University of South Bihar, Gaya-824236 (Bihar). India*
[6]*Japan Synchrotron Radiation Research Institute, Sayo 679-5198, Japan*
[7]*Department of Applied Physics, Tokyo University of Science, Katsushika-ku, Tokyo 125-8585, Japan*
[8]*Photon Factory, IMSS, High Energy Accelerator Research Organization, Tsukuba, Ibaraki 305-0801, Japan*
[9]*Department of Applied Physics, Waseda University, Shinjuku-ku, Tokyo 169-8555, Japan*

## Abstract

The prototypical diluted magnetic semiconductor $Cd_{1-x}Mn_x Te$ is a spin glass ($x<0.6$) or an antiferromagnet ($x>0.6$), but becomes ferromagnetic upon doping with a small amount of Cr atoms substituting for Mn. In order to investigate the origin of the ferromagnetism in $Cd_{1-x-y}Mn_x Cr_y Te$, we have studied its element specific magnetic properties by x-ray absorption spectroscopy (XAS) and x-ray magnetic circular dichroism (XMCD) at the Cr and Mn $L_{2,3}$ edges. Thin films were grown by molecular beam epitaxy with a fixed Mn content of $x \sim 0.2$ and varying Cr content in the range of $y = 0 – 0.04$. Measured XAS and XMCD spectra indicate that both Cr and Mn atoms are divalent and that the ferromagnetic or superparamagnetic components of Cr and Mn are aligned in the same directions. The magnetization of Mn increases with increasing Cr content. These results can be explained if ferromagnetic interaction exists between neighboring Mn and Cr ions although interaction between Mn atoms is largely antiferromagnetic. We conclude that each ferromagnetic or superparamagnetic cluster consists of ferromagnetically coupled several Cr and a much larger number of Mn ions.

## Introduction

II-VI semiconductor compounds doped with a high concentration of Mn atoms such as $Cd_{1-x}Mn_xTe$, $Zn_{1-x}Mn_xTe$, and $Hg_{1-x}Mn_xTe$ were first synthesized in 1980's and were called diluted magnetic semiconductors (DMS's) [1-7]. They have attracted attention because of their unique magnetic and optical properties such as Faraday rotation, spin-glass behavior, and the formation of magnetic polarons. Galazka *et al.* [3] and Nagata *et al.* [4] have investigated the magnetic properties of Mn-doped CdTe and HgTe, respectively, and found paramagnetic, spin-glass, and antiferromagnetic phases as a function of temperature and Mn content. Mn-doped CdTe, $Cd_{1-x}Mn_xTe$ (CdMnTe), exhibits a giant Faraday rotation under a magnetic field and is used as an optical isolator in magneto-optical devices [5-7]. It has been believed that the interesting magnetic and magneto-optical properties arise from *sp-d* interaction between the magnetic ions and the band electrons and/or interaction between the magnetic ions themselves [8,9].

Shen *et al.* [10] reported that Cr doping into $Cd_{1-x}Mn_xTe$ turns the system from the antiferromagnetic spin glass to a ferromagnet. They studied the ferromagnetic behavior of $Cd_{1-x-y}Mn_xCr_yTe$ bulk crystals by magnetization measurements. Interaction between the Cr magnetic moments are considered to be ferromagnetic because Cr-doped II-VI semiconductors are ferromagnetic [11,12]. In order to understand the origin of the ferromagnetism in Cr-doped CdMnTe, evaluating the nature of interaction between the Mn and Cr magnetic moments is crucial. For this purpose, x-ray absorption spectroscopy (XAS) and x-ray magnetic circular dichroism (XMCD) are ideal tools because they are element specific probes of the electronic and magnetic properties of transition-metal ions in solids. In the present work, therefore, we have studied the electronic and magnetic properties of the Mn and Cr ions in $Cd_{1-x-y}Mn_xCr_yTe$ thin films by XAS and XMCD experiments.

## Experimental

Thin films of $Cd_{1-x-y}Mn_xCr_yTe$ were fabricated on GaAs (001) substrates by the molecular beam epitaxy (MBE) method. The Cr content $y$ was varied from 0 to 0.04 while keeping the Mn content $x$ fixed at ~0.20. First, a buffer layer of CdTe of ~ 700 nm thickness was grown on the GaAs (001) substrate and after that a $Cd_{1-x-y}Mn_xCr_yTe$ layer of ~ 300 nm thickness was successively grown on the CdTe buffer layer. The sample surface was capped with a 2 nm thick Al layer to avoid surface oxidization. X-ray diffraction studies confirmed that the $Cd_{1-x-y}Mn_xCr_yTe$ films were grown epitaxially without the formation of any secondary phases.

Mn and Cr $L_{2,3}$-edge ($2p$ core-to-$3d$) XAS and XMCD measurements were done at the undulator beam line BL-16 of Photon Factory (PF), KEK. XAS and XMCD spectra were collected in the total-fluorescence yield (TFY) mode using a photodiode. The probing depth of the fluorescence yield detection was about 100 nm. A fixed magnetic field was applied perpendicular to the film surface while measuring XAS spectra, and the photon helicity was reversed to obtain the XMCD spectra. All the measurements were performed at 15 K under magnetic fields up to 5 T.

## Results and discussion

Figure 1 shows the Mn $L_{2,3}$-edge (Mn $2p$-$3d$) XAS and XMCD spectra of a $Cd_{0.8-y}Mn_{0.2}Cr_yTe$ thin film with $y$=0.04 taken in the TFY mode at T = 15K. Here, $\mu_+$ and $\mu_-$ refer to absorption spectra for photon helicity parallel and antiparallel to the Mn $3d$ spin, respectively. In Fig. 1(a), peaks around $hv$ = 643.46 and 653.96 eV are due to absorption from the Mn $2p_{3/2}$ (Mn $L_3$) and $2p_{1/2}$ (Mn $L_2$) core levels to the Mn $3d$ levels and their line shapes closely resemble those of the Mn $L_{2,3}$-edge XAS of $Ga_{1-x}Mn_xN$, where Mn is in the 2+ state and tetrahedrally coordinated by four N atoms [13]. Therefore, we conclude that the doped Mn ions in the $Cd_{1-x-y}Mn_xCr_yTe$ thin films are in the $Mn^{2+}$ state tetrahedrally coordinated by four Te atoms. Figure 1(b) shows the Mn $L_{2,3}$-edge XMCD spectra of the $Cd_{0.8-y}Mn_{0.2}Cr_yTe$ ($y$=0.04) sample measured at various magnetic fields. It is shown that the line shape of the XMCD spectrum does not change with magnetic field, but the intensity increases with it.

Figure 2 shows the Cr $L_{2,3}$-edge XAS and XMCD spectra of the $Cd_{0.8-y}Mn_{0.2}Cr_yTe$ ($y$=0.04) thin film. Peaks in the spectrum around $hv$ = 579.4 and 589.1 eV are due to absorption from the Cr $2p_{3/2}$ (Cr $L_3$) and Cr $2p_{1/2}$ (Cr $L_2$) core levels to the Cr $3d$ levels, respectively. The Cr $L_{2,3}$-edge XAS and XMCD spectra resemble those of $Zn_{1-x}Cr_xTe$ and show multiplet structures, suggesting that Cr takes the valence of 2+ with localized $3d$ electrons [14]. The Cr $L_{2,3}$ absorption overlaps the tail of broad absorption due to Te $3d$ → Te $5p$ transition (Te $M_{4,5}$ edge). The XMCD ($\mu_+$-$\mu_-$) spectra for different magnetic fields are shown in Fig. 2(b). Comparison of Figs. 1(b) and 2(b) reveals that the sign of Mn $L_{2,3}$ XMCD is the same as that of Cr $L_{2,3}$ XMCD, indicating that the magnetization of Mn and that of Cr are parallel to each other.

By applying the XMCD sum rules [15, 16] to the present data, we have estimated the magnetization of the Mn and Cr atoms separately. Because the application of the XMCD sum rules to TFY data does not necessarily give accurate numbers because of non-negligible self-absorption effect, the deduced magnetization due to the Mn and Cr ions should be taken as approximate and relative ones. Nevertheless, the self-absorption may not be so serious in the present samples because of the dilute concentrations of the Mn and Cr atoms [17]. Figure 3 shows the magnetization of the Cr and Mn 3$d$ electrons in the Cd$_{0.8-y}$Mn$_{0.2}$Cr$_y$Te ($y$=0.04) thin film as a function of magnetic field (divided by temperature, $H/T$, where $T$ = 15 K). The Cr and Mn magnetization as functions of magnetic field linearly increases at low magnetic fields and the slope becomes smaller at high magnetic fields, which enables us to decompose each of the Cr and Mn magnetization into the paramagnetic (PM) and superparamagnetic (SPM) components. (Here, we consider that the SPM behavior arises from nano-scale ferromagnetic (FM) clusters and, therefore, their signals are referred to as the FM/SPM components.) The PM component of magnetization linearly increases with magnetic field and the FM/SPM component of magnetization is given by a Langevin function L(ζ) [18-20].

$M_{Mn} = f_{Mn} m_{Mn}$ L($\mu\mu_0 H/k_B T$) + (1-$f_{Mn}$)$C_{Mn}$ $\mu_0$ $H/(T- \Theta_{P,Mn})$,  (1)

$M_{Cr} = f_{Cr} m_{Cr}$ L($\mu\mu_0 H/k_B T$) + (1-$f_{Cr}$)$C_{Cr}$ $\mu_0$ $H/(T- \Theta_{P,Cr})$,  (2)

L($\xi$) = coth($\xi$) – 1/$\xi$ ,  (3)

where $f_{Mn(Cr)}$ (0 ≦ $f_{Mn(Cr)}$ ≦ 1) denotes the fraction of Mn (Cr) atoms participating in the FM/SPM components, $m_{Mn(Cr)}$ the net magnetic moment of the Mn (Cr) atoms in the FM/SPM clusters, $\mu$ the total magnetic moment per FM/SPM cluster, $C_{Mn(Cr)}$ the Curie constant of Mn$^{2+}$ (Cr$^{2+}$) ions, and $\Theta_{P,Mn(Cr)}$ the Weiss temperature. Note that the parameters $f$, $m$, $C$, and $\Theta_P$ can be different for each element while $\mu$ is common.

The sum of the FM/SPM and PM components fitted to the magnetization data is shown by solid curves in Fig. 3. Because the XMCD line shapes do not change with magnetic field, both the FM/SPM and PM components of the Cr and Mn magnetization seem to originate from the same kinds of Cr$^{2+}$ and Mn$^{2+}$ ions, respectively. Figure 3 shows that the FM/SPM component per ion is larger for the Cr ion than the Mn ion, implying that the Cr ions drive the FM/SPM behavior of Cd$_{1-x-y}$Mn$_x$Cr$_y$Te and that the FM/SPM component of the Mn ion is induced by interaction with the spins of Cr atoms.

In order to determine the fitting parameters in Eq. (1), we have assumed the Weiss temperatures ($\Theta_P$) of the paramagnetic Mn$^{2+}$ and Cr$^{2+}$ ions to be -20 K [3], and added the constraint of 0 ≦ $f_{Mn}$, $f_{Cr}$ ≦ 1. We find from the fitting that most of Cr atoms contribute to the FM/SPM components ($f_{Cr}$ = 1.0 ± 0.1) while only 50% of Mn atoms are in the FM/SPM states ($f_{Mn}$ = 0.52 ± 0.05). The total magnetic moment of a FM/SPM cluster $\mu$ was 68 ± 6 $\mu_B$. The magnetic moments of Cr and Mn in the

FM/SPM clusters ($m_{Cr}$ and $m_{Mn}$) were 1.6 ± 0.1 $\mu_B$ and 0.76 ± 0.04 $\mu_B$, respectively. Note that the fitted values of $f_{Cr}$, $m_{Cr}$ and $\mu$ remain almost unchanged when the Weiss temperature $\Theta_P$ is varied from 0 K to -40 K, but $f_{Mn}$ changes from 0.80 ± 0.02 to 0.25 ± 0.1, $m_{Mn}$ from 0.50 ± 0.04 to 1.6 ± 0.3 $\mu_B$. Meanwhile, the product of $f_{Mn}$ and $m_{Mn}$, which represents the saturation magnetization of Mn atoms, stayed constant (0.4 $\mu_B$ for Mn). The Weiss temperatures smaller than -40 K did not reproduce the data well. Despite the above-mentioned uncertainties, it can be said that the Cr atoms induces the FM/SPM states because majority of the Cr atoms are described by the Langevin function while considerable amount of Mn atoms remain paramagnetic. Since the finite amount of Mn atoms shows FM/SPM behavior, there should be ferromagnetic interaction between Cr and Mn atoms. Each FM/SPM cluster has the magnetic moment of ~70 $\mu_B$.

Figure 4(a) shows the Cr concentration dependence of the Mn 2$p$ XMCD spectra of $Cd_{0.8-y}Mn_{0.2}Cr_yTe$ samples. The spectra indicate that the Mn $L_{2,3}$ XMCD intensity increases with Cr concentration. Figure 4(b) shows the Cr concentration dependence of the magnetization of the Mn ion as a function of magnetic field. From this figure, it is clear that Mn ions without Cr show a PM behavior as reported by Galazka *et al*. [3] and Nagata *et al*. [4], while after incorporating a small amount of Cr ions ($y$=0.04: 20% of the Mn ions), the FM/SPM component appears. A similar trend has also been observed in the magnetization measurements by Shen *et al*. [10] and by Ishikawa and Kuroda [21]. Ishikawa and Kuroda observed a negative Weiss temperature $\Theta_P$ for $Cd_{1-x}Mn_xTe$ but the $\Theta_P$ turned to positive after incorporating only 0.46 % of Cr. Therefore, a small amount of Cr present in the CdMnTe matrix changes $\Theta_P$ from a negative to a positive value. According to the present results, the apparent sign change of $\Theta_P$ may be due to the superposed FM/SPM component on the PM component.

In order to present an intuitive picture of the magnetic properties of $Cd_{1-x-y}Mn_xCr_yTe$, we show in Fig. 5 a schematic drawing of the spin configurations in the Cr-doped CdMnTe arising from the dominantly antiferromagnetic Mn-Mn interaction and the ferromagnetic Mn-Cr interaction as discussed above. In CdMnTe, the Mn spins are mutually antiferromagnetically coupled but the coupling is not so strong and the Mn spins can be aligned parallel in an applied magnetic field of ~0.4 T [5]. When a small amount of Cr is doped, the spins of Mn ions near Cr ions are canted towards the direction of the Cr spin, resulting in the Mn net magnetization parallel to that of Cr. As mentioned above, the saturation magnetization of Mn and Cr ions in a FM/SPM cluster was estimated to be 0.4 $\mu_B$ and 1.6 $\mu_B$, respectively, which are 8% and 40% of that of the fully spin-polarized Mn and Cr ions for $x$=0.2 and $y$=0.04. Therefore, the net numbers of Mn and Cr ions contributing to the FM/SPM component of magnetization are nearly the same. This combined with the total magnetic moment of a FM/SPM cluster of about 70 $\mu_B$ implies that a FM/SPM Cr-Mn complex consists of several $Cr^{2+}$ ions surrounding by a larger number of partially polarized $Mn^{2+}$ ions having a similar net polarization as that of the $Cr^{2+}$ ions. Such a FM/SPM cluster schematically shown in Fig. 5 acts as a building block of the FM/SPM behavior.

As for the origin of the ferromagnetic interaction between the $Mn^{2+}$ and $Cr^{2+}$ ions, Shen *et al*. [10] proposed a bound magnetic polaron mechanism, that is, a small fraction of $Cr^{2+}$ ions become $Cr^+$ and a hole bound to the $Cr^+$ acceptor mediates FM coupling between the Mn ions contained within the Bohr radius of the hole. However, the energy level of the $Cr^+$ charge state has been estimated 1.34 eV above the valence-band maximum [22], which is too deeply located within the band gap to be populated at low temperatures. Another possible origin of the FM/SPM cluster formation is FM superexchange interaction between the Cr and Mn ions because, according to the Goodenough-Kanamori rule, neighboring $Mn^{2+}$ (with the $d^5$ configuration) and $Cr^{2+}$ (with the $d^4$ configuration) tend to be coupled ferromagnetically. In the $x$=0.2 and $y$=0.04 sample, each Cr ion has a few Mn ions at the nearest-neighbor sites, and the average distance between Cr ions is at most only twice as large as that between Mn ions. Therefore, the Mn-Cr FM coupling can connect several Cr atoms through the Cr-Mn-Cr bonds to form a FM/SPM cluster. In order to confirm this scenario, more detailed XMCD studies including temperature-dependent measurements are desired.

## Conclusion

We have performed XAS and XMCD measurements of $Cd_{1-x-y}Mn_xCr_yTe$ with the Mn content of $x \sim 0.2$ and the Cr content varying in the range of $y = 0 – 0.04$. The valences of Cr and Mn atoms were found to be 2+. The magnetization of the Mn ions increases with increasing Cr content, indicating that in the presence of the $Cr^{2+}$ ions the spins of $Mn^{2+}$ ions tend to be aligned ferromagnetically with Cr. We found that each FM/SPM cluster contain several Cr ions surrounded by a larger number of Mn ions with the total magnetization similar to that of the Cr ions.

**Acknowledgments**

We would like to thank Kenta Amemiya and Masako Sakamaki for valuable technical support. This work was supported by Grants-in-Aid for Scientific Research (15H02109, 15K17696, and 19K03741) from JSPS, the Quantum Beam Technology Development Program from JST and SERB-DST (ECR/2016/001741). The experiment

**Data Availability**

The data that support the findings of this study are available from the corresponding author upon reasonable request.

*Corresponding author: shoya.sakamoto@issp.u-tokyo.ac.jp

magnetic semiconductor (Cd,Mn,Cr)Te, AIP Conf. Proc. **1399**, 705 (2011).

[22] P. Moravec, M. Hage-Ali, L. Chibani and P. Siffert, Deep levels in semi-insulating CdTe, Mater. Sci. Eng. B **16**, 223 (1993).

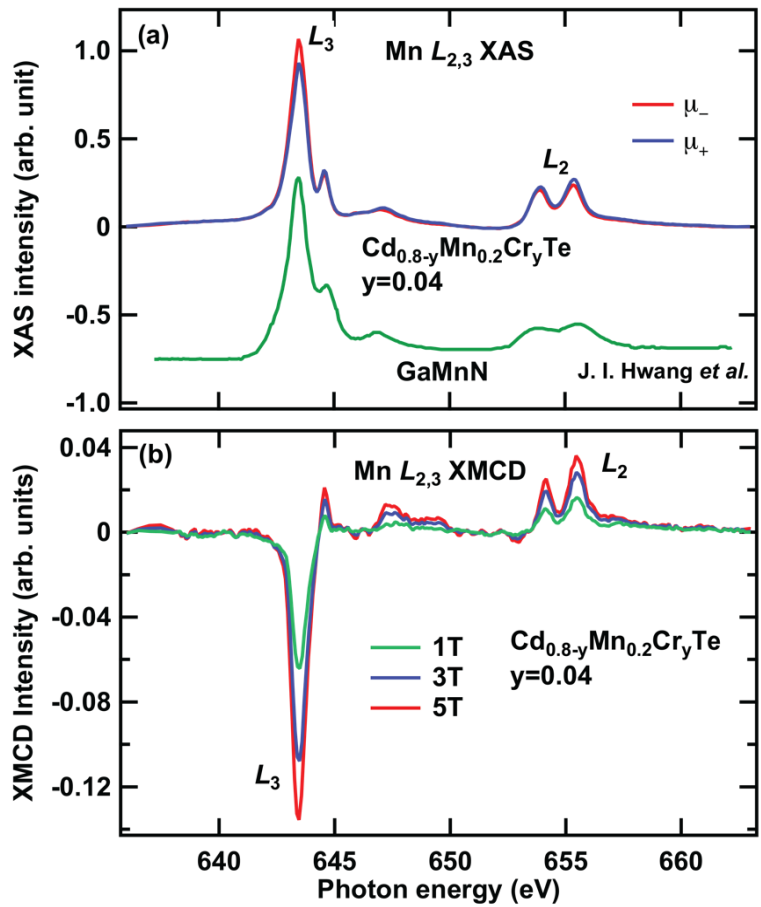

FIG. 1: Mn $L_{2,3}$-edge XAS and XMCD spectra of $Cd_{0.8-y}Mn_{0.2}Cr_yTe$ thin film with $y=0.04$. (a) XAS spectra compared with that of $Ga_{1-x}Mn_xN$ (with $Mn^{2+}$) [13], Copyright (2005) by the American Physical Society. (b) magnetic field dependence of XMCD spectra.

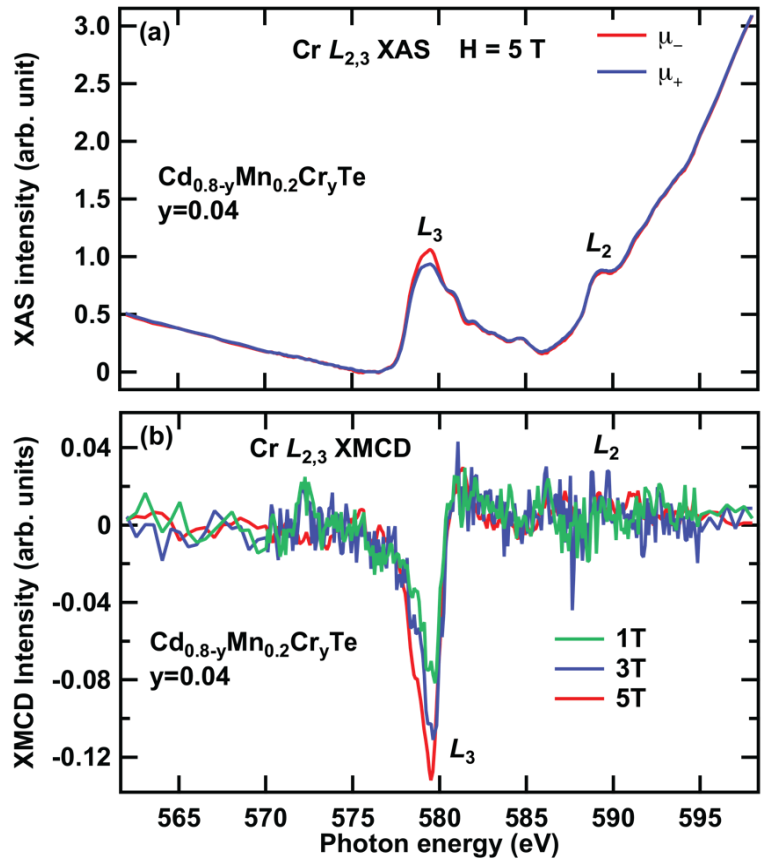

FIG. 2: Cr $L_{2,3}$-edge XAS and XMCD spectra of $Cd_{0.8-y}Mn_{0.2}Cr_yTe$ thin film with $y=0.04$. (a) XAS spectra. The background above ~580 eV is due to the tail of Te $M_{4,5}$ XAS. (b) Magnetic field dependence of XMCD spectra.

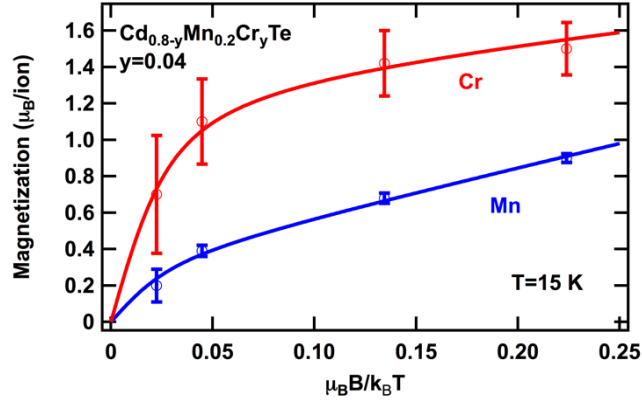

FIG. 3: Element dependence of the magnetization curve of the Mn and Cr atoms deduced from the $L_{2,3}$ XMCD spectra using the XMCD sum rules. The data have been fitted to a Langevin function representing the FM/SPM component plus a linear-$H$ function representing the paramagnetic component as given by Eqs. (1) and (2). The magnetization of Mn and that of Cr can be explained if several Cr and a larger number of Mn ions are ferromagnetically coupled in each FM/SPM cluster.

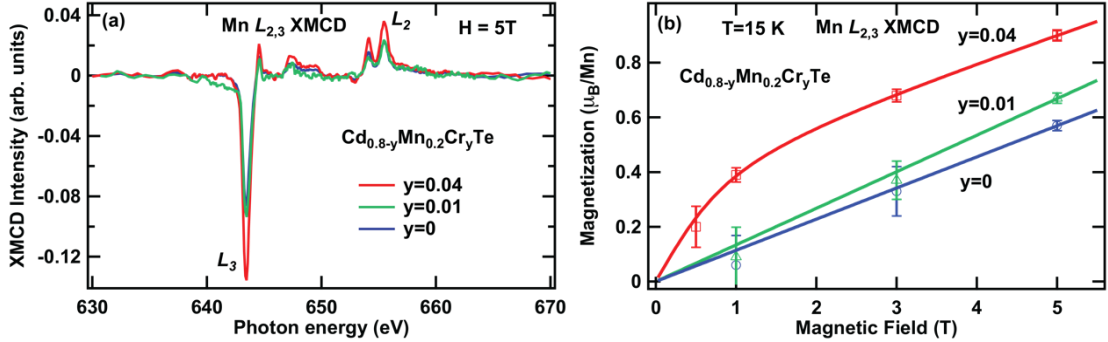

FIG. 4: (a) Cr concentration dependence of Mn $L_{2,3}$ XMCD spectra of $Cd_{0.8-y}Mn_{0.2}Cr_yTe$ thin films. (b) Cr concentration dependence of magnetization curve of the Mn atom deduced from the Mn $L_{2,3}$ XMCD spectra. The data have been fitted to a convex combination of Langevin function representing the FM/SPM component and a linear-$H$ function representing the paramagnetic component [Eqs. (1) and (2)].

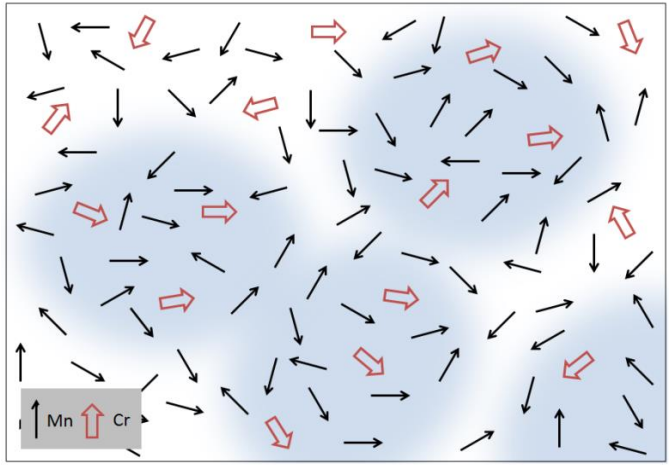

FIG. 5: Schematic picture of the $Mn^{2+}$ (thin black arrows) and $Cr^{2+}$ (thick white arrows) ions in Cr-doped $Cd_{1-x}Mn_xTe$. $Mn^{2+}$ ions are antiferromagnetically coupled mutually, but those surrounding $Mn^{2+}$ ions are partially polarized parallel to the $Cr^{2+}$ ions and this polarization cloud forms FM/SPM cluster.